\documentclass[a4paper]{spie}  

\usepackage{amsmath,amsfonts,amssymb}
\usepackage{graphicx}
\usepackage[colorlinks=true, allcolors=blue]{hyperref}
\usepackage{bm}
\usepackage{subfig}

\usepackage{bbding}
\usepackage{pifont}
\usepackage{amssymb} 

\title{Visualizing intestines for diagnostic assistance of ileus based on intestinal region segmentation from 3D CT images}

\author[a]{\ \ Hirohisa Oda}
\author[b]{Kohei Nishio}
\author[c]{Takayuki Kitasaka}
\author[d,a]{Hizuru Amano}
\author[a]{Aitaro Takimoto}
\author[a]{Hiroo Uchida}
\author[e]{Kojiro Suzuki}
\author[b]{Hayato Itoh}
\author[b]{Masahiro Oda}
\author[b,f,g]{Kensaku Mori}

\affil[a]{Nagoya University Graduate School of Medicine, Japan} 
\affil[b]{Graduate School of Informatics, Nagoya University, Japan} 
\affil[c]{School of Information Science, Aichi Institute of Technology, Japan} 
\affil[d]{Graduate School of Medicine, University of Tokyo, Japan} 
\affil[e]{Department of Radiology, Aichi Medical University, Japan} 
\affil[f]{Information Technology Center, Nagoya University, Japan} 
\affil[g]{Research Center for Medical Bigdata, National Institute of Informatics, Japan} 

\authorinfo{Further author information: (Send correspondence to H. Oda and K. Mori)\\
H. Oda : E-mail: hoda @ mori.m.is.nagoya-u.ac.jp, Telephone: +81 52 789 5688\\
K. Mori: E-mail: kensaku @ is.nagoya-u.ac.jp, Telephone: +81 52 789 5689\\ \\
Copyright 2020 Society of Photo-Optical Instrumentation Engineers (SPIE)}

\begin{document} 
\maketitle

\begin{abstract}
This paper presents a visualization method of intestine (the small and large intestines) regions and their stenosed parts caused by ileus from CT volumes. Since it is difficult for non-expert clinicians to find stenosed parts, the intestine and its stenosed parts should be visualized intuitively. Furthermore, the intestine regions of ileus cases are quite hard to be segmented. The proposed method segments intestine regions by 3D FCN (3D U-Net). Intestine regions are quite difficult to be segmented in ileus cases since the inside the intestine is filled with fluids. These fluids have similar intensities with intestinal wall on 3D CT volumes. We segment the intestine regions by using 3D U-Net trained by a weak annotation approach. Weak-annotation makes possible to train the 3D U-Net with small manually-traced label images of the intestine. This avoids us to prepare many annotation labels of the intestine that has long and winding shape. Each intestine segment is volume-rendered and colored based on the distance from its endpoint in volume rendering. Stenosed parts (disjoint points of an intestine segment) can be easily identified on such visualization. In the experiments, we showed that stenosed parts were intuitively visualized as endpoints of segmented regions, which are colored by red or blue.
\end{abstract}


\begin{figure}[t]
\centering
\subfloat[FCN structure: 3D U-Net]{
\centering
 \includegraphics[clip, trim=0 14cm 0 0, page=3, height=4.2cm]{illusts.pdf}
}\\
\subfloat[Patch cropping]{
\centering
 \includegraphics[clip, trim=0 12cm 4cm 0, page=1, height=5cm]{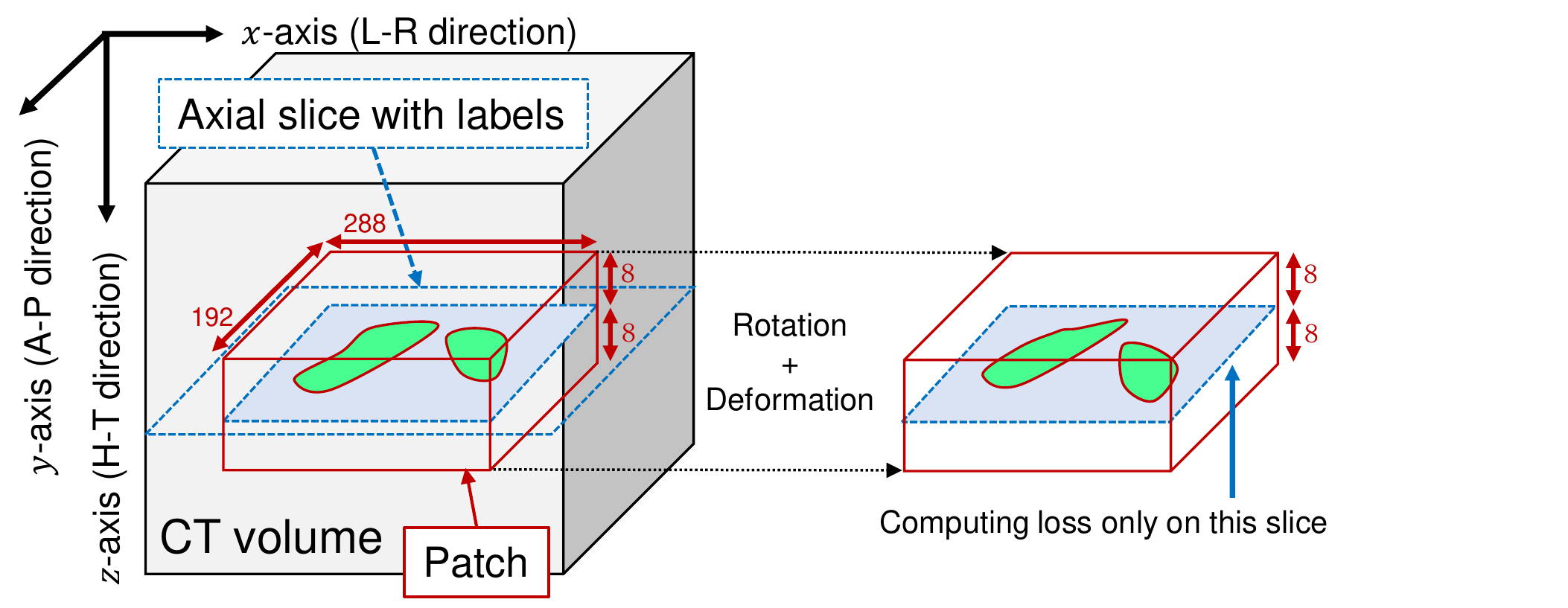}
}
\caption{Network and patches. (a) Structure of 3D U-Net. (b) Patch cropping. Axial slice having labels and their neighboring slices are obtained as patch consisting of $(288, 192, 16)$ voxels. \label{fig_training}}
\end{figure}

\section{Introduction}

In this paper, we propose a visualization method of the intestine (the small and large intestines) and their stenosed parts caused by ileus on CT volumes. On diagnosis of ileus, clinicians manually find obstruction or stenosed parts of the intestines by tracking the path of intestines on volumetric CT images (CT volumes). However, finding stenosed parts on CT volumes is difficult for non-expert clinicians since the small intestine is long and winding. Visualization of intestines and their stenosed parts assists clinicians to understand how the intestines are running and identify the stenosed parts. 

There are various visualization methods of the large intestine (colon) for CT volumes. The virtual colonoscopy \cite{lefere2010virtual,ganeshan2013virtual} generates colonoscopy-like images. By exploring inside the large intestine as if using the colonoscopy, users can find lesions such as polyps. Virtual unfolding of the large intestine \cite{halier2000nondistorting,yao2010reversible} also have been widely studied. Entire the large intestine can be observed all at one as the unfolded view. Since intensities of feces or liquid inside the large intestine are similar to their surrounding tissues, fecal tagging is desired for clear visualization of the colon. Patients orally administrate contrast agent to increase visibility of feces or fluid on CT volumes. Virtual cleansing of feces \cite{serlie2010electronic,tachibana2018deep} are for replacing fecal or fluid regions by intensities of the air. It makes the entire large intestine as if filled with the air.

However, fecal tagging is not possible for diagnosis of ileus, which is often required on emergency diagnosis. Furthermore, visualization methods mentioned above are only for the large intestine. The ileus often occur in the small intestine, which is much more long and have complicated shape than the large intesitine. Accurate and intuitive visualization method specific for ileus diagnosis is desired.

Intestine segmentation is a fundamental process to visualize the intestine region for diagnostic assistance of the ileus case. There are very few segmentation methods that can be applied to the small intestines. Zhang, et al. \cite{zhang2013mesenteric} segmented small intestines on contrast-enhanced CT angiography (CTA) scans. Most intestinal segmentation methods \cite{franaszek2006hybrid,yang2014multilabel} are applied only for the large intestines with assumption of fecal tagging by contrast agents. Virtual cleansing of feces \cite{serlie2010electronic,tachibana2018deep} mentioned in the previous paragraph is also assuming fecal tagging. In summary, compared to segmentation of large intestines with fecal tagging, segmentation for ileus patients are difficult due to 1) low contrast between fluid and intestinal wall and 2) long and winding shape of the small intestines.

Utilization of 3D FCNs, such as 3D U-Net \cite{cciccek20163d}, would be one of solutions for segmenting intestine regions of ileus cases from CT volumes. However, it is very difficult to prepare 3D manual traced labels of the intestine including not only the large intestine but also the small intestine. Alternatively, we introduce weak annotation approach using small sample data. Training is possible by manually traced data of the intestine region the intestine only on several axial slices.
Note that this work is not semi-automated segmentation with sparse annotation \cite{cciccek20163d,sugino2018automatic}, which produces entire segmentation results from volumetric images that partly contains manual tracing. Our proposed method does not require any manual tracing on testing data.

3D U-Net generally produces some false positive (FP) regions. We implement manual selection interface to select intestine segments (connected components) from 3D U-Net output for eliminating those FP regions. We select points inside intestinal regions for eliminating FP regions. Endpoints of an intestinal segment are then identified by utilizing constrained distance transformation. Each segment is colored based on distance from an endpoint to assist a clinician easily to understand running status of an intestinal segment. Also, this makes easy to find stenosed parts, which are disjoint of intestinal segments.

\section{Methods}

\subsection{Overview \label{sec_overview}}

The proposed method performs visualization of stenosed parts by segmentation of intestines by 1) intestinal segmentation and 2) endpoint detection of intestinal segments. Training of the 3D U-Net is required in prior to inference. We assume that CT volumes for both training and inference are of ileus patients. In the following procedure, we utilize isotropic volumetric images that can be obtained by interpolation process of original CT volumes.

Weak annotations are manually created on several (7 in our experiments) axial slices of each CT volume in the training dataset.  Those axial slices are randomly selected from CT volumes by a computer scientest who knows ileus diagnosis well. 

\subsection{Training \label{sec_network}}


Our network structure is designed based on the 3D U-Net\cite{cciccek20163d} as illustrated in Fig. \ref{fig_training}(a). On behalf of implementing skip connections as concatenation of features, we use summation \cite{roth2018towards}. This allows us to reduce the number of parameters to be trained (green arrows in Fig. \ref{fig_training}(a)) compared to concatenation. 

In training, we fix a patch size as $288 \times 288 \times 16$. Since patch size along $z$-axis is small (16 voxels), padding techniques inserting fixed values (e.g. zero-padding) affect central part of feature maps after several times of convolutions. Therefore, padding around boundaries of feature maps is performed by refraction before each convolutional layers to keep the feature map size. Size of network output is the same as the input patch. The output represent probabilities that each point is outside (near from 0) or inside (near from 1) the intestines.

As illustrated in Fig. \ref{fig_training}(b), patches are cropped so that they contain manually-traced ground-truth as $8^{\text{th}}$ slice of each patch. We compute the loss function only at the $8^{\text{th}}$ slice of a patch volume. For more robust training using small amount of dataset, we apply non-rigid deformation and random rotation for each slice of patch volumes. For non-rigid deformation, we define a $g \times g$-grid on a patch. Each grid point randomly moves $d$ pixels for each of $x$- and $y$-axes. Random rotation is performed for $r$ degrees in maximum.

\subsection{Intestinal segmentation}

Intestinal regions are segmented by using the trained 3D U-Net by weak annotation approach. Only several slices on each CT volumes in the training dataset should have manually-traced labels. The sizes of interpolated volumes along with $x$- and $y$-axes are adjusted to become the next highest multiple of 32 to fit to the FCN by padding.

Patch volume is cropped from the interpolated input volume and then is fed to
the 3D U-Net to obtain segmentation results (Fig. \ref{fig_training} (b)).
We use middle slice ($8^{\text{th}}$ slice) of inference results of the patch volume. We obtain inference results for all of slices of the interpolated volume by cropping a patch volume in the sliding window manner
 in one voxel stride to $z$-axis direction. When we input one patch volume to the FCN, we use only the inference result on
the middle slice ($8^{\text{th}}$) slice. This is because the FCN training is performed by the
loss function computed only at the middle slice of the patch volume. By iterating this process, we obtain segmentation results for whole volume.

Trained FCN can take any sizes of patch volume of $(32n_1, 32n_2, 16) \: (n_1, n_2
\in I)$ unless exceedance of GPU memory limit is caused. Since FCN consists of convolution and
the training process learns convolution kernel parameters, we can implement FCN
that can change input patch volume size in the inference process.

\subsection{Visulizing stenosed parts as endpoints of intestinal segments\label{sec_seg}}

3D U-Net still produces some FP regions that needs to be removed. Wrong connections between neighboring intestinal regions are also produced in segmentation results. To remove such regions, firstly morphological opening operation is performed to 3D U-Net outputs for eliminating small connected component and departing adjacent intestine regions. 

\begin{figure}[tb]
\centering
\includegraphics[clip, trim=0 3cm 0 0, page=4, height=8cm]{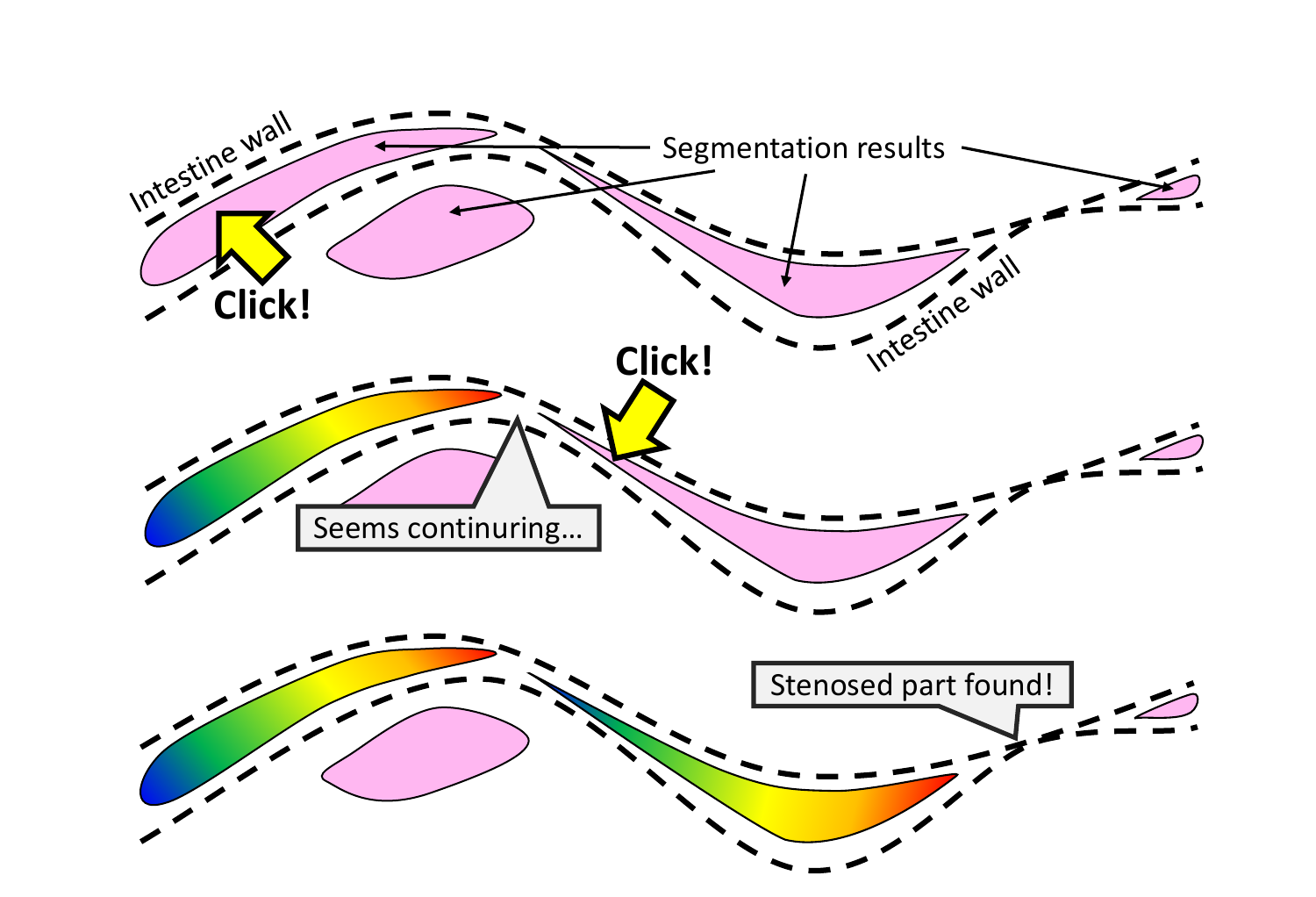}
\caption{Concept of our manual selection interface. By clicking one of segmetation results inside the intestines, the result is colored for visualizing intestinal segment. One of endpoints becomes blue and the other becomes red. If subsequent intestinal segment exists, users can click again it. Clicking for several times allows users to reach stenosed part. \label{fig_interface}}
\end{figure}

\begin{figure}[tb]
\centering
\includegraphics[clip, trim=0 9.5cm 1cm 0, page=6, width=16cm]{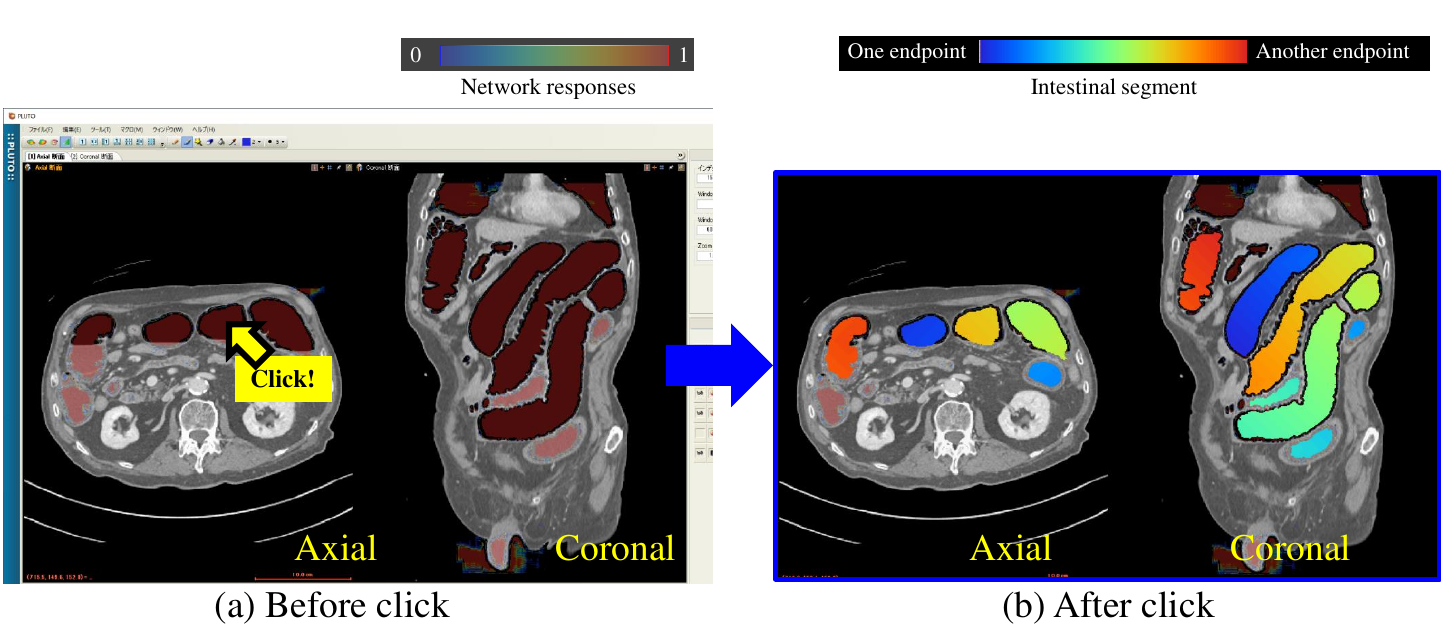}
\caption{Screenshots of manual selection interface. (a) Before click: Network responses are shown as semitransparent heatmap. Users click inside intestines (high responses of network). (b) After click. Clicked intestinal segment is shown clearly. One endpoint is blue and another is red. \label{fig_screenshot}}
\end{figure}

As the concept illustrated in Fig. \ref{fig_interface}, we developed a manual selection interface of intestinal segment. User just need to click somewhere inside the intestine (no need to click either end of the intestinal segment). The intestinal segment including the clicked point is obtained. Screenshots are shown in Fig. \ref{fig_screenshot}

\subsection{3D visualization}

We visualize intestinal segments by using 3D volume rendering. Each segment is colored based on the distance from an endpoint along the running path of each intestinal segment. Constrained distance transformation is utilized to find endpoints of an intestinal segment. One point is selected in an intestinal segment and then constrained distance transformation is performed again from the selected point. We consider a voxel that has the maximum distance value as the one of endpoints. Then we perform a constrained distance transformation from an endpoint computed in the previous step. The voxel that has the maximum distance value is considered as another endpoint. An intestinal segment is colored by a distance from the endpoint firstly computed in volume rendering of intestinal segments.

\begin{figure}[tb]
\centering
 \includegraphics[clip, trim=0 8.8cm 10cm 0, width=10cm, page=1]{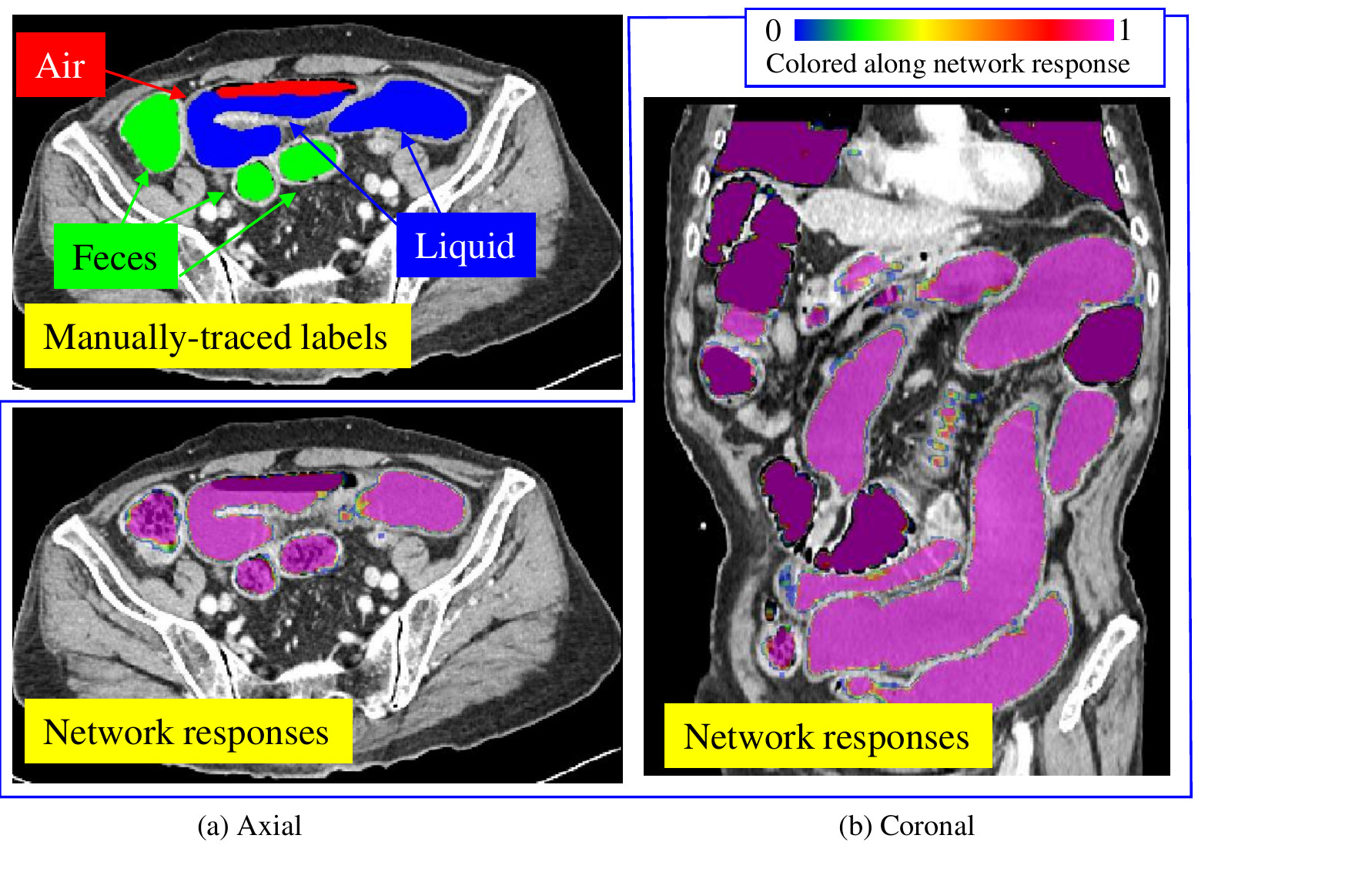}\\
 \includegraphics[clip, trim=0 8.8cm 14cm 0, width=10cm, page=2]{screenC.pdf}
\caption{Case C whose stenosed part was successfully visualized. \label{fig_example}}
\end{figure}

\section{Experiments}

Seven CT volumes of ileus patients were allowed to use under IRB approval of Aichi Medical University (Aichi, Japan). They consists of $512 \times 512 \times (260-316)$ voxels with resolution $ (0.625-0.789) \times (0.625-0.789) \times 2.0 \text{mm}^3$. Seven axial slices had manually-traced labels of the intestine contents, which were checked by a pediatric surgeon. Stenosed parts are also found by the surgeon on all CT volumes. Those CT volumes were separated into three groups (three groups consist of 2, 2 and 3 CT volumes, respectively) for cross-validation.

Mini-batch size $m$ was set as four due to memory limitation (24GB) of the Quadro P6000 (NVIDIA) GPU. Training was continued for 20000 iterations using Adam optimizer with initial training rate $10^{-3}$. The network was implemented using the Keras. Segmentation accuracies are quantitatively calculated on axial slices having manually-traced labels by 1) Dice score for just after thresholding the generated probabilities and 2) average number of connections between neighboring intestinal segments. Other parameters are set as follows: $g=5$, $d$=10 pixels and $r=20$ degrees.

\section{Results}

Quantitative evaluation results for all cases are shown in Table \ref{table_eval}. Mean Dice score among 7 cases was 0.75, representing accurate segmentation was performed for almost all cases.

Figure \ref{fig_example} shows results of Case C. A long intestinal segment was segmented and visualized very well under high Dice score (0.84). In contrast, Case A shown in Fig. \ref{fig_screenA} visualized only a very short intestinal segment. Dice score of Case A was 0.53, which is the worst case among the 7 cases. 


 \section{Discussions}
 
 Case C shown in Fig. \ref{fig_example} has relatively clearer contrast than other cases. Network responses were high in almost the entire intestine regions. Although wrongly produced high responses were produced inside the lungs (Fig. \ref{fig_example} (b)), these FPs were successfully removed from final visualization results (Fig. \ref{fig_example}(c)) thanks to our manual selection interface illustrated in Fig. \ref{fig_interface}. The stenosed part caused by ileus was successfully visualized as one of endpoints colored in blue. Correct coloring was done since intestinal segments were properly segmented without connections.
 
\begin{figure}[tb]
\centering
 \includegraphics[clip, trim=0 0 15cm 0, page=1, width=12cm]{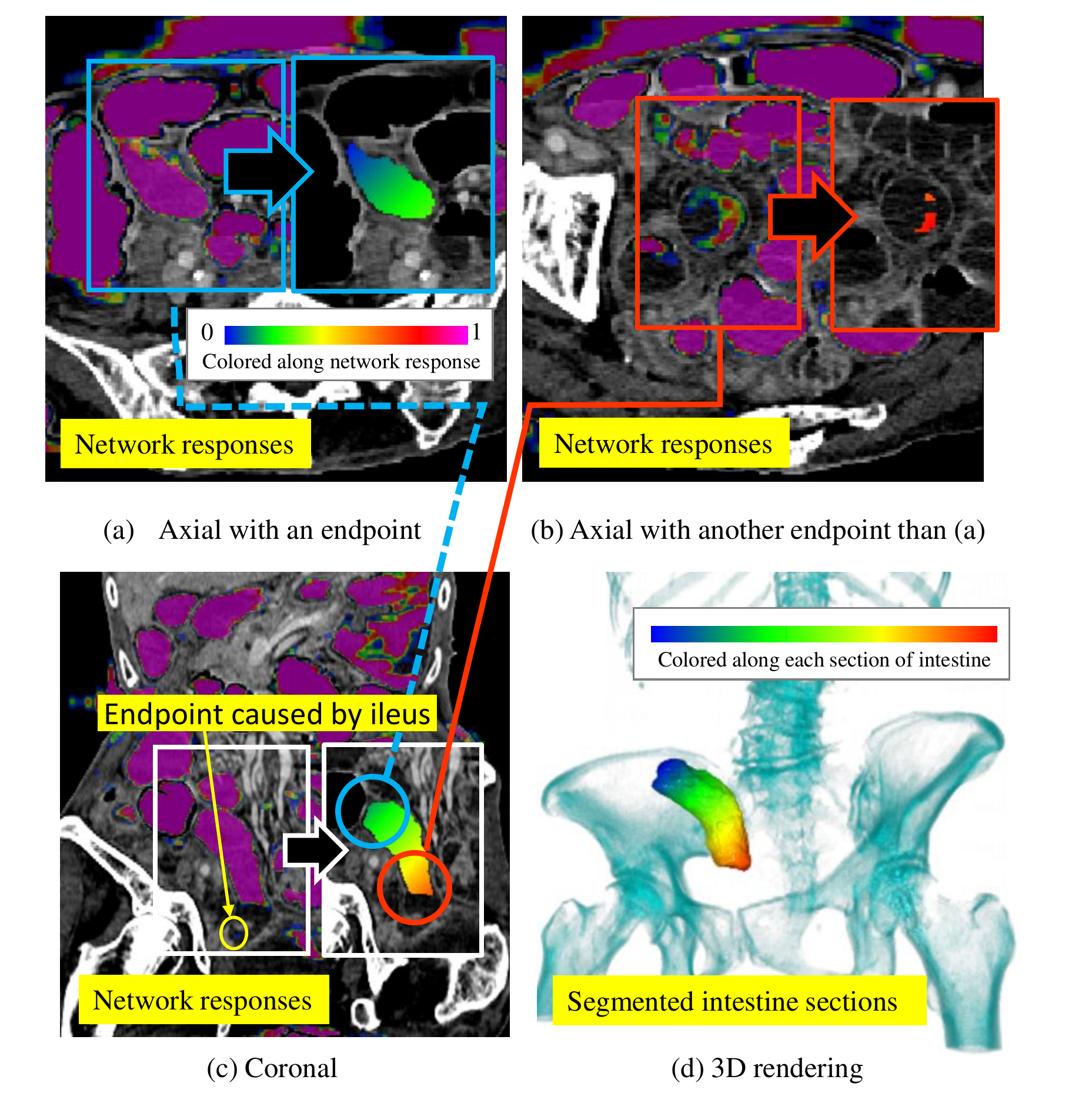}
\caption{Results of Case A that segmentation was not performed until stenosed part. \label{fig_screenA}}
\end{figure}

 Case A shown in Fig. \ref{fig_screenA} failed to reach the stenosed part due to false negatives of 3D U-Net. Intestine walls have relatively low contrast between intestine walls and surrounding tissues than other cases. More robust network structure to low contrast images than current one should be considered in the future.

\begin{table}
\caption{Quantitative evaluation: Dice score (higher is better) and  mean number of connections (lower is better).  \label{table_eval} }
\centering
\begin{tabular}{cc|cc}
\hline
Case & Cause of ileus & Dice & \# connections  \\
\hline
A & Obturator hernia & 0.53 & 0.14    \\
B & Colon cancer & 0.67 & 1.00       \\
C & Postoperative adhesion (suspicion) & 0.84 & 0.29    \\
D & Unknown & 0.86 & 0.00   \\
E & Paralysis (suspicion)& 0.88 & 0.29   \\
F & Sigmoidal volvulus & 0.58 & 0.33   \\
G & Strangulate &  0.88 &  0.00  \\
\hline
Mean & & 0.75 & 0.29   \\
\hline
\end{tabular}
\end{table}


\section{Conclusions}

We proposed a visualization method for assistance of diagnosing ileus. Segmentation of intestinal segments is based on training with weak annotation approach. Manual selection interface of intestinal segment removes false positives. Furthermore, the click-based interface well received by clinicians which assist them to reach stenosed parts intuitively, easily and quickly. Intestinal segmentation was successfully done using 3D U-Net with weak annotation approach. Either end of an intestinal segment is visualized by our coloring scheme. Future work includes tuning network structure for intestine regions. Also, it is required to find a way of quantitative evaluation of stenosed part detection.

\bibliography{spie2020_hoda} 
\bibliographystyle{spiebib} 

\end{document}